\documentclass[twocolumn]{aastex631}
\usepackage{amsmath}
\usepackage{wasysym}
\usepackage{graphicx}
\usepackage{amssymb}
\usepackage{epstopdf}
\usepackage{mathrsfs}
\usepackage{anyfontsize}
\usepackage{natbib}
\usepackage{CJK}

\usepackage{color}
\usepackage{lipsum}
\usepackage{diagbox}
\usepackage{gensymb}
\usepackage{booktabs}
\usepackage{threeparttable}
\usepackage[figuresright]{rotating}
\usepackage{multirow}

\shorttitle{}
\shortauthors{Hu et al.}
\begin{document}
	\begin{CJK*}{UTF8}{gbsn}

	\title{Investigating lower limit of metallicity for Galactic thin disk}

    \correspondingauthor{Zhengyi Shao}
	\email{zyshao@shao.ac.cn}
		
	\author[0000-0003-1828-5318]{Guozhen Hu (胡国真)}
 \altaffiliation{Physics Postdoctoral Research Station at Hebei Normal University}
 \affiliation{Department of Physics, Hebei Normal University, Shijiazhuang 050024, Peopleʼs Republic of China}
 \affiliation{Physics Postdoctoral Research Station at Hebei Normal University, Shijiazhuang 050024, Peopleʼs Republic of China}
	\affiliation{Shanghai Astronomical Observatory, Chinese Academy of Sciences, 80 Nandan Road, Shanghai 200030, China}

	\author[0000-0001-8611-2465]{Zhengyi Shao (邵正义)}
	\affiliation{Shanghai Astronomical Observatory, Chinese Academy of Sciences, 80 Nandan Road, Shanghai 200030, China}
    \affiliation{Key Lab for Astrophysics, Shanghai 200234, China}

\author{Erbil G\"{u}gercino\u{g}lu}
	\affiliation{National Astronomical Observatories, Chinese Academy of Sciences, 20A Datun Road, Chaoyang District, Beijing 100101, China}
 
\author[0000-0003-1359-9908]{Wenyuan Cui (崔文元)}
	\affiliation{Department of Physics, Hebei Normal University, Shijiazhuang 050024, Peopleʼs Republic of China}

\begin{abstract}

We explore the metal-poor regime of the Galactic disk on the distribution of stars in the [$\alpha$/M]-$V_{\phi}$ plane, to identify the most metal-poor thin disk (MPTnD) stars belonging to the low-$\alpha$ sequence. Chemical abundances and velocities of sample stars are either taken or derived from APOGEE DR17 and Gaia DR3 catalogs. We find the existence of a well-separated extension of the kinematically thin disk stars in the metallicity range of -1.2 $<$[M/H]$<$ -0.8 dex. Based on two-by-two distributions of [Mg/Mn], [Al/Fe] and [C+N/Fe], we further confirmed 56 high-possibility metal-poor thin disk (HP-MPTnD) giant stars and suggested the lower metallicity limit of the thin disk below -0.95 dex. A comparative analysis of HP-MPTnD sample with other Galactic components revealed its chemo-dynamical similarities with canonical thin disk stars. These low-$\alpha$ metal-poor stars are predominantly located in the outer disk region and formed in the early stage of the formation of thin disk. Their existence provides compelling support for the two-infall model of the Milky way's disk formation. Moreover, these stars impose observational constraints on the timing and metallicity of the second gas infall event.

\end{abstract}
	

\section{Introduction}\label{sec:intro}

It is widely accepted that the Milky Way (MW)'s disk consists of two prominent components, the thick and thin disks, with different geometrical, kinematic and chemical features (e.g., \citealt{1983MNRAS.202.1025G,2014A&A...567A...5R,2015A&A...583A..91G,2016MNRAS.461.4246W,2017A&A...608L...1H,2018MNRAS.481.1645M}). Compared to the thick disk, the thin disk has a higher rotational velocity and a lower velocity dispersion (e.g., \citealt{2015A&A...583A..91G,2016A&A...586A..39R, 2020MNRAS.494.3880B, 2020A&A...643A.106B}). With the accumulating of high-resolution spectroscopic data from large survey projects, the disk stars are found to exhibit bimodality in the $\alpha$-enhancement ([$\alpha$/M]) vs. metallicity ([M/H]) plane (e.g., \citealt{2011A&A...535L..11A,2014A&A...562A..71B,2015ApJ...808..132H}). This bimodality is then used to distinguish the chemically defined Galactic populations (e.g., \citealt{2011A&A...535L..11A,2014A&A...562A..71B}). The high-$\alpha$ sequence is recognized as the thick disk, whereas the low-$\alpha$ sequence is mainly associated with the thin disk. This bimodal pattern provides important constraints regarding the formation and evolution of the MW, such as the gas accretion history, the star formation efficiency and the gas outflow process (e.g., \citealt{2017ApJ...835..224A,2017ApJ...837..183W,2023ApJ...950..142H}). Nevertheless, the formation scenario of the Galactic disk remains controversial, in particular, the origin of the chemical bimodality still remains to be understood.
\\

Two alternative formation models for the MW disk stars have been proposed to reproduce the observed bimodality in the chemical abundance plane. One is the so-called two-infall model, which suggests that the thick and thin disks are distinct Galactic components formed in two separate phases of the gas infall \citep[e.g.,][]{Grisoni19,Grisoni20a,Grisoni20b,Spitoni19,Spitoni20,Lian20,Lian20b}. The earlier gas infall triggers a starburst, thereby accelerating the formation of thick disk stars \citep{Lian20b}. While the second gas accretion phase is mainly associated with the secular formation of thin disk stars \citep{Chiappini97,Lian20}. Another disk formation model suggests that thick and thin disks are \emph{``two parts of a single continuous disk component which evolves with time due to the continued gas accretion, star formation of thin disk stars and disk heating”} \citep{2020MNRAS.491.5435B,2021MNRAS.507.5882S,2021ApJS..254....2P}.  In this paper, we refer to it as a continuous-accretion model.\\

Although both models could qualitatively reproduce the observed $\alpha$-dichotomy, they may differ in their quantitative predictions for the number density distribution in the [$\alpha$/M]-[M/H] plane, in particular at the metal-poor tail of the low-$\alpha$ sequence. In the two-infall model, the metal-poor thin disk (MPTnD) stars were formed later from the interstellar medium diluted by the second infall accretion phase \citep{Chiappini97,Lian20}. As a result, these MPTnD stars would have lower [$\alpha$/M] ($<$ 0.2 dex), similar to the metal-rich thin disk stars. In the continuous accretion model, the metal-poor stars were formed in the disk at an earlier time. So even for the kinematically thin-disk like stars, their [$\alpha$/M] values would be higher ([$\alpha$/M] $>$ 0.25 dex) and comparable to those of the thick disk at similar metallicity. Therefore, searching the most MPTnD stars, namely investigating the lower limit of the metallicity for the thin disk, could help to discriminate the different disk formation scenarios. \\

Unfortunately, it is not an easy task to identify the most MPTnD  stars. Many works have proposed that there is a significant overlap between the MPTnD stars and the accreted halo stars in terms of chemical abundance (\citealt{Nissen10,Hawkins15}), which poses the identification in an ambiguous way.\\

Recently, increasing numbers of surveys and works are studying metal-poor stars of MW\citep[e.g.,][]{2020MNRAS.497L...7S,2020Natur.581..269N,2022ApJ...938L...2F,2023NatAs...7..611R,2023MNRAS.521.1045R,2024A&A...688A.167N}. Despite intense studies in the literature on the investigation of lower limit of metallicity for thin disk, no consensus has been achieved yet. In earlier works, the thin disk stars are generally thought to be have [Fe/H] $>$ -0.7 dex \citep[e.g.,][]{Reddy03,Fuhrmann04,Soubiran05}. \cite{Mishenina04} identified kinematically thin disk stars could be down to [Fe/H] $\sim$ -1.0 dex. \cite{2014A&A...562A..71B} separated 714 F and G dwarf stars into thick and thin disks by using their kinematic characteristics and confirmed that [Fe/H] $\approx$ -0.7 dex is the low limit of metallicity for thin disk stars. Another study by \cite{Hawkins15} suggested that the chemical and kinematic characteristics of stars are consistent with canonical thin disk stars if one extends to lower metallicity at [Fe/H] $\sim$ -0.85 dex. More recently, by means of kinematic methods,  \cite{Fern2021} proposed that the metallicity of the thin disk may be down to $\sim$ -2 dex. \\

Combine the kinematic data, mainly the rotational velocity $V_{\phi}$, and the $\alpha$-enhancement to investigate the MPTnD stars is of interest because both higher $V_{\phi}$ and lower [$\alpha$/M] are the signature features of thin disk stars. Thanks to precise chemical abundances and radial velocities provided by the high-resolution spectroscopy survey of APOGEE and astrometric measurements from the Gaia that enable us to effectively pinpoint MPTnD stars on the [$\alpha$/M]-$V_{\phi}$ plane. Furthermore, by utilizing additional abundances such as [Mg/Mn] and [Al/Fe], we can further refine our identification of MPTnD stars to distinguish them from accreted halo stars. \\

In our previous study (\citealt{2022ApJ...929...33H}, hereafter HS22), we proposed a metallicity threshold of [M/H]$\sim -0.8$ dex to differentiate between the thin disk and halo components when we used the Gaussian mixture model (GMM) to analysis different disk components. This demarcation was primarily due to the limited number of thin disk stars with [M/H] $<$ -0.8 dex, hindering their inclusion as a distinct component in the GMM fitting process. Given that the present work aims to identify the most metal-poor stars, we will extend our investigation below this threshold within the thin disk population.\\

The paper is organized as follows. In Section 2, we present the sample selection criteria and describe the observational or derived parameters of sample stars. In Section 3, we initially select MPTnD candidates based on the values of [$\alpha$/M] as a function of $V_{\phi}$ (Section~\ref{sec:results}). Then, we identify the high possibility (HP-) MPTnD stars based on the comparison of their [Mg/Mn], [Al/Fe] and [C+N/Fe] distributions (Section~\ref{3.2: chem}). We continue to compare the HP-MPTnD stars with other Galactic components in terms of spatial distribution, orbital parameters (Section~\ref{3.3}) and age distribution (Section~\ref{3.4}) to ensure that they belong to the tail of thin disk. The implications of the disk formation scenario are discussed in Section 4. Conclusions are presented in Section 5. \\

\section{Sample Selection and Data Reduction}\label{sec:data}

\begin{figure}[htbp] 
	\centering
	\includegraphics[width=0.5\textwidth]{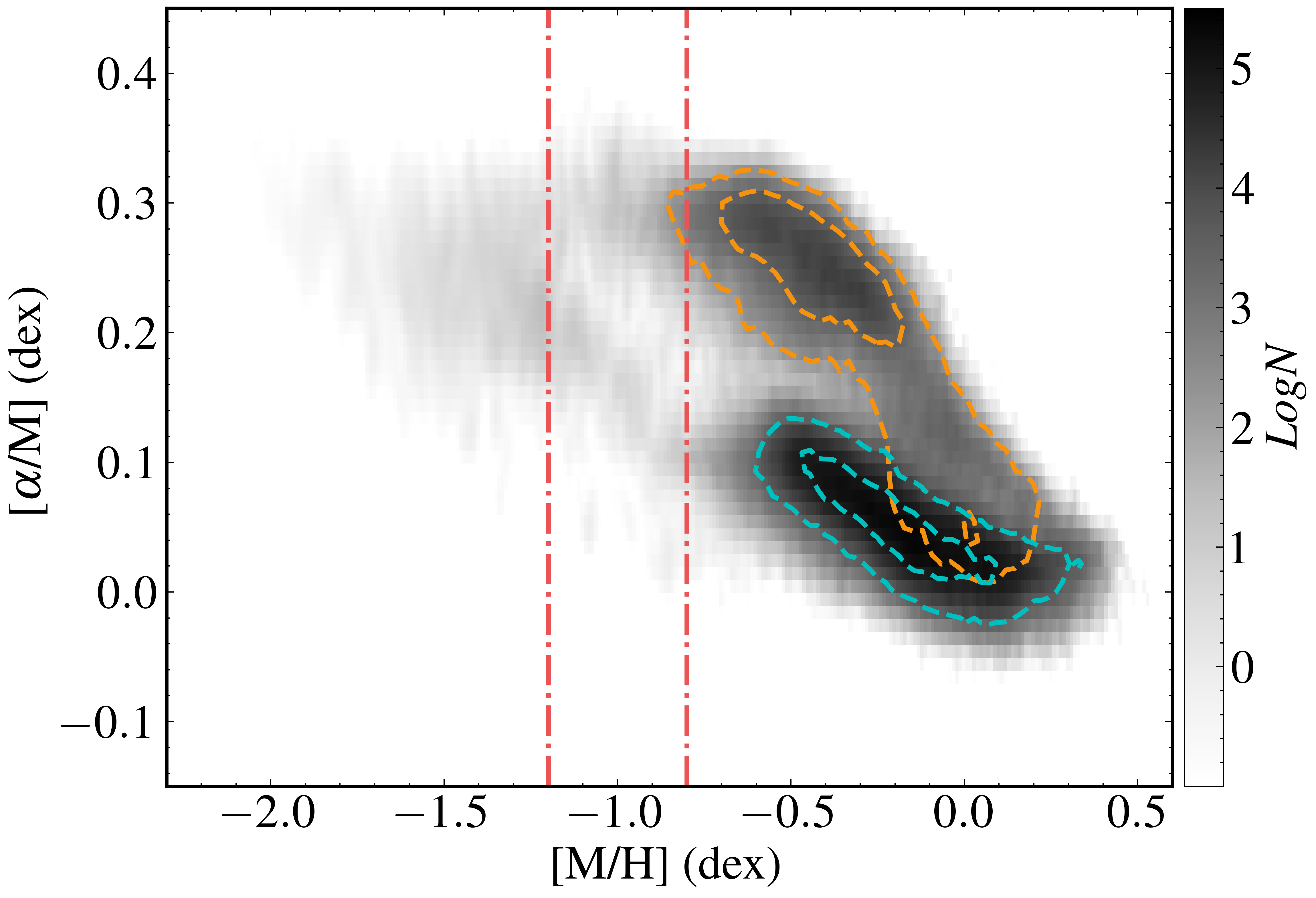} 
	\caption{The logarithmic greyscale plot of the number density of sample stars in the [$\alpha$/M]-[M/H] plane. As defined in HS22, the dashed orange and cyan lines correspond to the 1$\sigma$ and 2$\sigma$ contours for the high-$\alpha$ and low-$\alpha$ sequences, respectively. Two red dashed-dotted lines at [M/H]=-1.2 dex and -0.8 dex divide the sample into three metallicity intervals.}
	\label{Fig:1}
\end{figure}

\subsection{Sample selection}

Our sample was obtained by cross-matching APOGEE DR17  \citep{2022ApJS..259...35A} with Gaia DR3 \citep{2021A&A...649A...1G, 2021A&A...649A...2L}. A cone search with a radius of one arcsec yielded a common catalog of 613,424 objects.  The resulting sample contains stellar coordinates ($\alpha$, $\delta$) and proper motions ($\mu_{\alpha}$, $\mu_{\delta}$) from Gaia \citep[][]{2021A&A...649A...2L}, as well as radial velocities ($RV$) and chemical abundances derived from the APOGEE Stellar Parameter and Chemical Abundances Pipeline (ASPCAP, \citealt{2016AJ....151..144G}). Additionally, we employ the photo-geometric distance ($d$) estimated by \cite{2021AJ....161..147B}, which is derived from the Gaia parallax and photometry. This extensive dataset provides multi-dimensional phase-space information for our stellar sample. \\

To ensure the quality of stellar parameters and elemental abundances, stars were selected from the APOGEE catalog according to the following criteria: {\tt ASPCAP\_CHI2} $<$ 10, {\tt ASPCAP\_FLAG} $=$  0 and {\tt SNR} $>$ 70. Additionally, stars located in fields targeting the bulge, known star clusters, and dwarf galaxies were excluded. These selection criteria resulted in the removal of approximately 60$\%$ of the stars from the original dataset.\\

For the Gaia data, to ensure reliable astrometric solutions, we selected stars with {\tt epsi} $<$ 4 and {\tt sepsi} $<$ 8, following the criteria of \cite[][]{2021ApJ...908L...5S}. To mitigate the impact binaries on astrometric estimates, we retained targets with a renormalized unit weight error ({\tt RUWE}) $<$ 1.4, as recommended by Gaia Collaboration \citep[][]{2021A&A...649A...2L}. Additionally, to construct an accurate map of stars in position and velocity space, we only included stars with a relative distance error  $\sigma_d$/$d < $ 10$\%$.\\

Finally, 128,532 stars are left as our parent sample. There are 119,752 giant stars ({\tt ASPCAP\_GRID} $=$ g ) and 8,780 dwarf stars ({\tt ASPCAP\_GRID} $=$ d) in our sample. Fig.~\ref{Fig:1} illustrates the distribution of these stars in the [$\alpha$/M] - [M/H] chemical abundance plane. It can be easily seen that there is a significant bimodal sequence of [$\alpha$/M] in the relatively metal-rich regime of [M/H] $>$ -1.2 dex. Moreover, at [M/H] $>$ -0.8 dex, the low-$\alpha$ sequence definitely consists of thin disk stars. On the other side, it has been suggested that at [M/H] $<$ -1.2 dex, the stars are predominantly halo stars (\citealt{Hawkins15}, HS22).  Keeping these in mind, we will focus on the stars within -1.2 $<$ [M/H] $<$ -0.8 dex (2,228 stars) in the present work, aiming to determine whether there are thin disk stars in the quoted metallicity range.\\

\subsection{Data Reduction}

The stellar chemical information are all taken directly from APOGEE DR17, such as [M/H], [$\alpha$/M], [Mg/Mn], [Al/Fe], [C/Fe], and [N/Fe], which will be discussed in following sections. The corresponding median uncertainties, $\sigma_{\rm [M/H]}$, $\sigma_{\rm [\alpha/M]}$, $\sigma_{\rm [Mg/Mn]}$, $\sigma_{\rm [Al/Fe]}$, $\sigma_{\rm [C/Fe]}$ and $\sigma_{\rm [N/Fe]}$ are 0.01, 0.01, 0.01, 0.03, 0.02 and 0.03 dex respectively. \\

Using the python package $galpy$ \citep{2015ApJS..216...29B}, we transformed coordinates ($\alpha$, $\delta$), proper motions ($\mu_{\alpha}$, $\mu_{\delta}$), distances ($d$) and radial velocities ($RV$) into the Galactocentric cylindrical frame $(R_{gc}, \phi, Z, V_{R}, V_{\phi}, V_{Z})$. We adopted solar Galactocentric coordinates ($R_\odot$, $Z_\odot$) = (8.125, 0.021) kpc \citep{2018A&A...615L..15G,2019MNRAS.482.1417B} and solar motion of (11.1, 242, 7.25) km\,s$^{-1}$ in radial, azimuthal, and vertical directions, respectively (\citealt{2010MNRAS.403.1829S,2019MNRAS.482.1417B}). Thus, the Galactic rotational velocities ($V_{\phi}$) were obtained, and their uncertainties were estimated by employing the Monte Carlo sampling method assuming a multivariate Gaussian distribution of their original data errors. Notably, the uncertainty in rotational velocity ($\sigma_{V_{\phi}}$) is about 2 km s$^{-1}$. \\

Additionally, we used observational parameters ($\alpha$, $\delta$, $\mu_{\alpha}$, $\mu_{\delta}$, $d$, $RV$) as initial conditions and utilized the GALPOT package to integrate stellar orbits of sample stars in the Galactic potential model of \cite{2017MNRAS.465...76M}. We calculated the pericenter ($r_{\rm peri}$) and apocenter ($r_{\rm apo}$) radii to derive the orbital eccentricity $ecc = (r_{\rm apo}-r_{\rm peri})/(r_{\rm apo} + r_{\rm peri})$ as well as the maximum vertical distance $Z_{\rm max}$ and the guiding radius $R_{\rm g}$ \footnote{$R_{\rm g}$ is defined by $R_{\rm g}$ = $R_{\rm gc} V_{\phi}$/$V_{c}$, with $V_{c}$ $=$ 230 km s$^{-1}$}. We also calculated stellar orbital inclination ($\theta_L$), i.e. the angle between the stellar orbit and the MW plane. \\

\section{ Searching and identifying metal-poor thin disk stars} \label{sec:3}

\subsection{ A low-$\alpha$ and fast-rotating overdensity in the metal-poor regime}\label{sec:results}

\begin{figure*}[htbp] 
	\centering
	\includegraphics[width=1\textwidth]{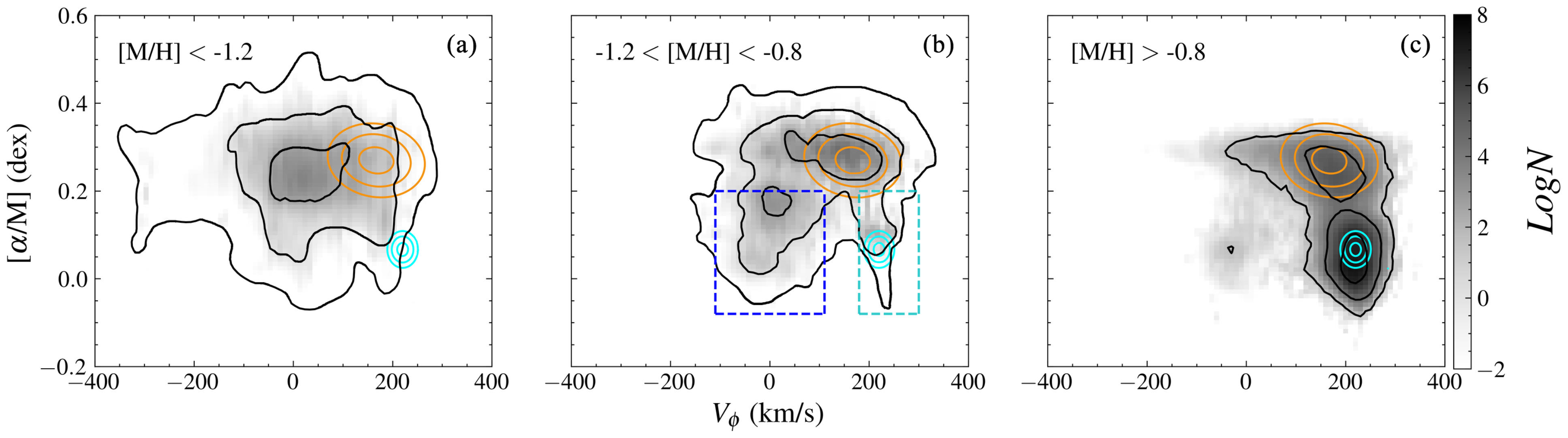} 
	\caption{The [$\alpha$/M]-$V_{\phi}$ distribution for the parent sample stars: (a) [M/H] $<$ -1.2 dex; (b) -1.2 $<$ [M/H] $<$ -0.8 dex; (c) [M/H] $>$ -0.8 dex. The black contours denote 1$\sigma$, 2$\sigma$ and 3$\sigma$ number distribution densities in each panel. The orange and cyan concentric ellipses represent the modelled distribution of thick and thin disk, respectively. In panel (b), the cyan and blue boxes delineate the location of selected the MPTnD star candidates and accreted halo stars, respectively. }
	\label{Fig:2}
\end{figure*}

Fig.~\ref{Fig:2} displays the smoothed number density distributions of three metallicity intervals of stars ([M/H] $<$ -1.2 dex, -1.2 dex $<$ [M/H] $<$ -0.8 dex and [M/H] $>$ -0.8 dex) in the [$\alpha$/M] - $V_{\phi}$ plane. Specifically, in order to facilitate comparison, we show the modelled distribution of canonical thick disk stars (orange concentric ellipses) and thin disk stars (cyan concentric ellipses) in all three panels. Their positions and dispersions are derived from the corresponding members of Table 2 of HS22. \\

In the most metal-poor regime, [M/H] $<$ -1.2 dex, (panel (a) of Fig.~\ref{Fig:2}), we can find that there is only one over-density region which is centered on [$\alpha$/M] $\sim$ 0.25 dex and $V_\phi \gtrsim 0$ km s$^{-1}$. It is noteworthy that while this over-density has large dispersions on both [$\alpha$/M] and $V_\phi$, it correlates neither to the thick disk nor to the thin disk. Clearly, this subsample is dominated by halo stars, 
which is consistent with previous studies suggesting that stars with [M/H] $<$ -1.2 overwhelmingly belong to the halo (i.e.  \citealt{Hawkins15}).  \\

Panel (b) of Fig.~\ref{Fig:2} reveals three distinct components within the metal-poor interval of -1.2 dex $<$ [M/H] $<$ -0.8 dex. One component exhibits high-$\alpha$ and $V_{\phi} \sim$ 140 km s$^{-1}$, characteristics consistent with the thick disk. A second component, characterized by low-$\alpha$ and $V_{\phi} \sim$ 0 km s$^{-1}$, aligns with the properties of the accreted halo. Notably, within the low-$\alpha$ region, an additional overdensity emerges at the location similar to the thin disk, centered on $V_{\phi} \sim$ 210 km s$^{-1}$ and [$\alpha$/M]$\sim$ 0.1 dex, indicating the presence of thin disk stars in this metal-poor range. The $V_\phi$ dispersion of these stars are not significantly larger than that of the canonical thin disk, which implies that they belong to the kinematically cool population. This observation highlights the fact that while the accreted halo and thin disk may overlap in [$\alpha$/M], they exhibit distinct $V_{\phi}$ distributions, resulting in a clear bimodality in the low-$\alpha$ region of this panel.\\

Panel (c) of Fig.~\ref{Fig:2} demonstrates that, for stars with [M/H] $>$ -0.8 dex, almost all of them belong to the thick disk or the thin disk. Besides, there is a small overdensity with low-$\alpha$ and $V_{\phi} \sim$ 0 km s$^{-1}$ which corresponds to accreted halo stars. This result suggests that the accreted stars may extend to a more metal-rich regime \citep[e.g.,][]{Nissen10,2018Natur.563...85H,2019MNRAS.482.3426M}. The metal-rich tail of accreted halo stars is interesting but out of the scope of this work, and we will address this issue in a separate paper. \\

To summarise the findings of panel (b), kinematically thin disc stars occur in the more metal-poor regime than -0.8 dex. Crucially, their relatively lower [$\alpha$/M] values enable differentiation from thick disk stars. We propose selecting stars with comparatively low [$\alpha$/M] and higher rotational velocities in the [$\alpha$/M] - $V_{\phi}$ plane as MPTnD candidates. Specifically, stars with 180 km s$^{-1} < V_{\phi} < 300$ km s$^{-1}$ and -0.08 dex $<$ [$\alpha$/M] $<$ 0.2 dex are indicated by the cyan box in this panel. This selection process yielded a total of 91 MPTnD candidates.\\

\begin{figure}[htbp] 
	\centering
	\includegraphics[width=0.5\textwidth]{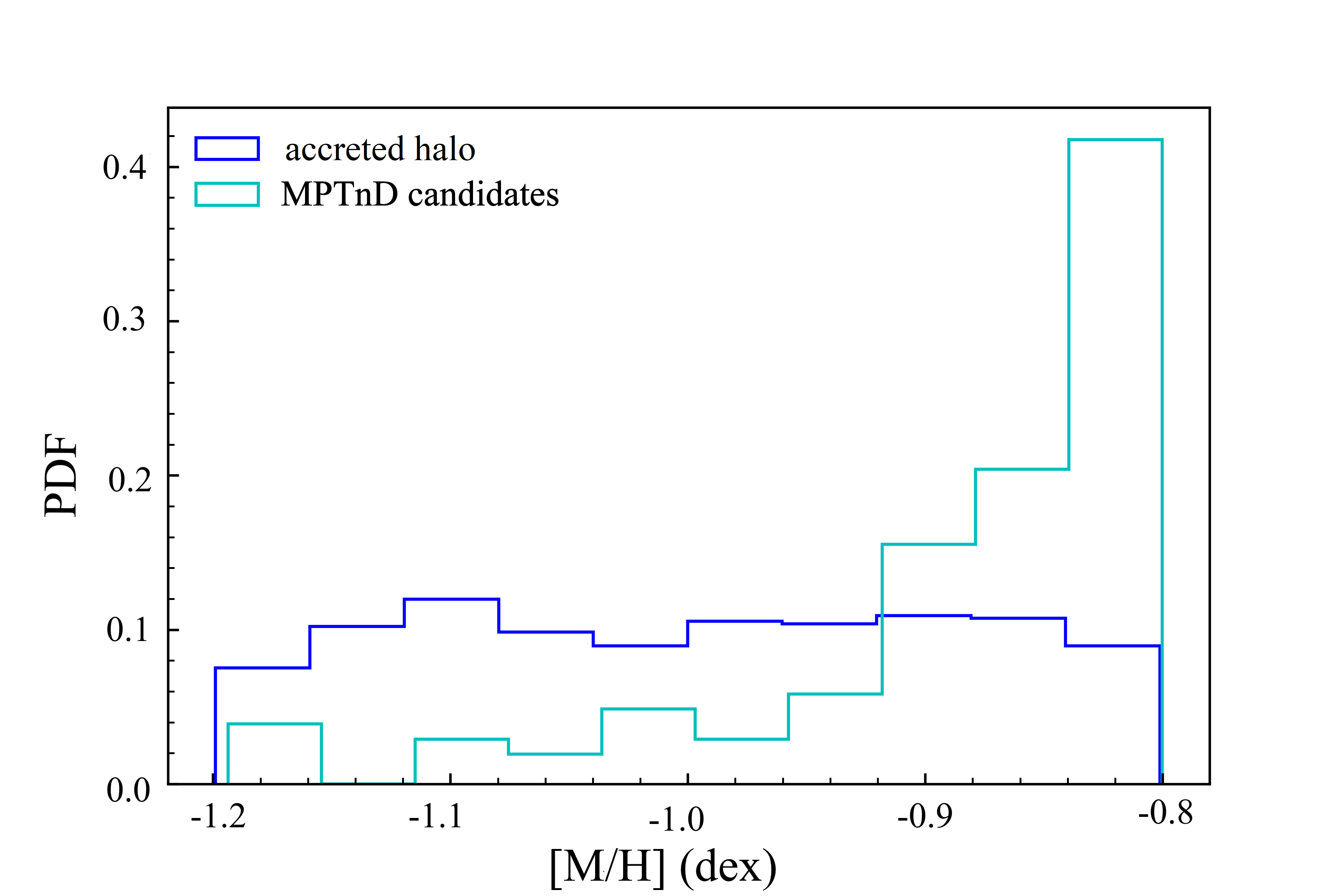} 
	\caption{The metallicity distribution of MPTnD candidates (stars in cyan box of Fig.~\ref{Fig:2}) and accreted halo stars (stars in blue box of Fig.~\ref{Fig:2}).}
	\label{Fig:select1}
\end{figure}

For comparison, we also select accreted halo stars by a blue box in panel (b) of Fig.~\ref{Fig:2}. The histograms in Fig.~\ref{Fig:select1} reveals the difference of the metallicity distributions of these two framed out populations. The accreted halo stars exhibit a uniform [M/H] distribution; while the MPTnD candidates, although present throughout the entire metal-poor range, are mainly concentrated in the region of [M/H] $>$ -0.95 dex. Anyway, it confirms that the MPTnD could be more metal-poor than -0.8 dex, and most likely, they may extend to -1.2 dex or even lower metallicity. \\

\begin{table}[h] 
\setlength{\abovecaptionskip}{0.05cm} 
\begin{threeparttable}  
\centering
\caption{Classification of MPTnD candidates within -1.2 $<$ [M/H] $<$ -0.8} \label{tab:1}

\begin{tabular*}{\hsize}{@{\extracolsep{\fill}}lrrr} 
\hline
\hline

Classification\tnote{a}& 
Num.\tnote{b}& 
[M/H]$<$-0.95\tnote{c}& 
[M/H]$>$-0.95\tnote{c}\\
\hline
HP-MPTnD stars& 56 & 7 (3) & 49 (29) \\
LP-MPTnD stars& 26 & 4 (3) & 22 (16)\\
accretion stars & 9 & 6 (0) & 3\, ( 0)\\
total & 91 & 17 (6)& 81 (45)\\
\hline

\hline
\end{tabular*}
 
\begin{tablenotes}
	\footnotesize
	\item[a] Classification of MPTnD candidates.
    \item[b] The number of stars.
    \item[c] The value in the parentheses is the number of stars with age measurements from \citet{2018MNRAS.481.4093S}.			 
\end{tablenotes}
\end{threeparttable}   
\end{table}

\subsection{Chemical Identification and Signature} \label{3.2: chem}

Our selection of MPTnD candidates likely suffers from contamination by accreted halo and thick disk stars due to their large dispersion in the [$\alpha$/M] - $V_\phi$ plane. To better distinguish between different stellar populations in the Galaxy, we can employ more specific element abundance ratios: [Mg/Mn], [Al/Fe], and [C+N/Fe] \citep[e.g.,][]{Hawkins15,Das20,Horta21,Carrillo22}. In order to avoid the bias of stellar element abundances obtained by ASPCAP when dealing with different types of stars \citep{2020ApJS..249....3A}, in this section, we only consider the giant stars (119,572 stars).\\

Magnesium (Mg) is the first element to be affected as a result of Type II Supernova (SNII) \citep{2013ARA&A..51..457N,Hawkins15} on a short timescale, and manganese (Mn) is a characteristic element of Type Ia Supernova (SNIa) enrichment on a much longer timescale. Therefore, [Mg/Mn] could be a better ``clock"  than [$\alpha$/M] to probe the star formation history of a given Galactic component \citep{Hawkins15}. Second, Aluminum (Al) is mainly produced by massive stars and dispersed in the ISM exclusively via SNII. So in general, [Al/Fe] ratio is expected lower for accreted systems \citep[e.g.,][]{Hawkins15,Hasselquist19,Hasselquist21}.  Moreover, because of the majority of giant stars are low-mass, where the Carbon plus Nitrogen ratio is conserved throughout the evolution of those stars, \cite{Hawkins15} suggested that the [C+N/Fe] also could distinguish Galactic components. All these three abundance ratios were obtained from APOGEE DR17. \\

\begin{figure*}[htbp] 
	\centering
	\includegraphics[width=1\textwidth]{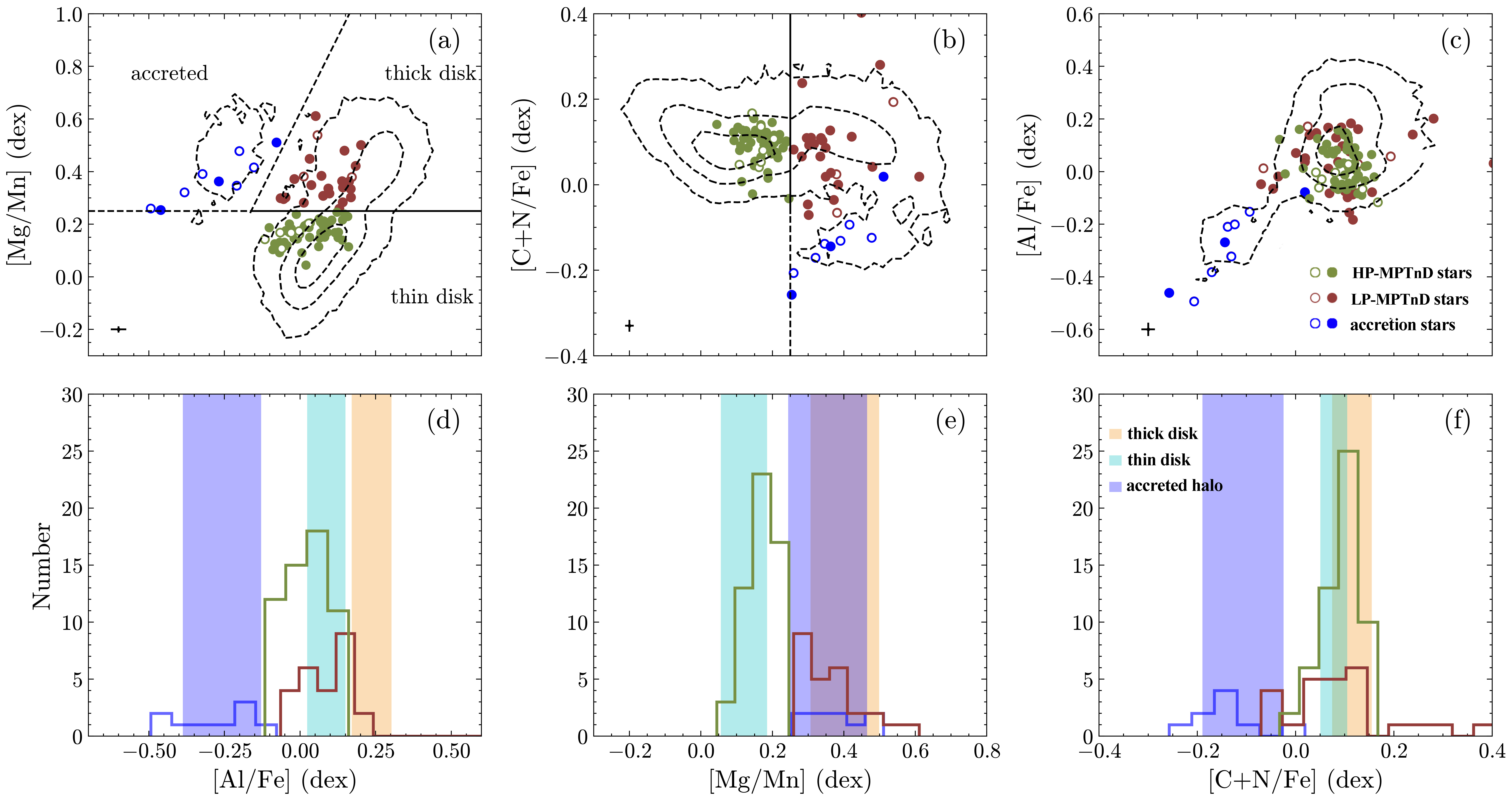} 
	\caption{Top panels from left to right: Projections of the MPTnD candidates along the [Mg/Mn]-[Al/Fe], [C+N/Fe]-[Mg/Mn] and [Al/Fe]-[C+N/Fe] planes. The error bars in the bottom-left corner of each panel represent the typical uncertainties of the corresponding parameters. The black contours denote 1$\sigma$, 2$\sigma$ and 3$\sigma$ number density distributions of the parent sample. In panel (a), the dashed line defines our criteria to separate in-situ disk and accretion star populations. The solid line further splits the disk into the thick disk and the thin disk. We also note that the quoted criteria is very similar to the one employed by \cite{Horta21,Naidu22}. The MPTnD candidates falling into the thin disk region are identified as HP-MPTnD stars shown in green circles, while stars falling into the thick disk region are classified as LP-MPTnD stars shown in brown circles, others are classified as accretion stars shown in blue circles. The stars with [M/H] $<$ -0.95 dex are shown with hollow circles. Bottom panels from left to right: The distributions of three classifications of MPTnD candidates in [Al/Fe], [Mg/Mn] and [C+N/Fe]. The orange, cyan, and purple shadows in the background are typical distribution ([16\%, 84\%]) of the thick disk, the thin disk and the accreted halo stars, respectively. The HP-MPTnD, LP-MPTnD stars and accretion stars are shown in the green, brown and blue histograms, respectively.}
	\label{Fig:chem1}
\end{figure*}

In top panels of Fig.~\ref{Fig:chem1}, MPTnD candidates (framed by the cyan box of Fig.~\ref{Fig:2}) are plotted on three different elemental abundance planes: [Mg/Mn]-[Al/Fe], [C+N/Fe]-[Mg/Mn], and [C+N/Fe]-[Al/Fe]. Black background contours delineate the number density distribution of the parent sample at 1$\sigma$, 2$\sigma$, and 3$\sigma$ levels, with an additional selection by using the criteria C/N/Mg/Al/Mn Fe FLAG = 0 from APOGEE. Following the classification scheme of \cite{Horta21} and \cite{Naidu22}, the [Mg/Mn]-[Al/Fe] plane (Fig.~\ref{Fig:chem1}a) is partitioned into three regions. Accreted halo stars occupy the area enclosed by dashed lines in the top left corner. Stars located to the right and below of these dashed lines represent in-situ disk stars. This in-situ population is further categorized into thick and thin disk stars based on [Mg/Mn] values separated by a horizontal solid line,  with lower values indicating thin disk stars. The figure employs distinct colors to differentiate MPTnD candidates into three populations: high-possibility metal-poor thin disk (HP-MPTnD) stars (green points), low-possibility metal-poor thin disk (LP-MPTnD) stars (brown points), and accretion stars (blue points). Among them, hollow circles mark stars with particularly low metallicity ([M/H] $<$ -0.95 dex). Table~\ref{tab:1} summarizes the numbers of MPTnD candidates across these classifications.\\

These three populations of MPTnD candidates can also be compared with known Galactic components through their histograms of [Al/Fe], [Mg/Mn], and [C+N/Fe] in lower panels of Fig.~\ref{Fig:chem1}. The distributions of accretion stars for all three elements match the distribution location of the general accreted halo (purple shadows). Therefore, they are surely rejected as MPTnD stars. HP-MPTnD stars almost follow the distribution of the canonical thin disk (green shadows), but with slightly lower [Al/Fe] and higher [Mg/Mn]. Comparing to the HP-MPTnD, the distributions of LP-MPTnD stars are more close to those of the thick disk (orange shadows). The most significant difference between HP-MPTnD and LP-MPTnD stars lies in their [Mg/Mn] distributions, which are consistent with the significant differences in the distribution of this element in the canonical thin and thick disks.  \\

In summary, through the analysis of three additional chemical elements, we have narrowed down our search and identified 56 candidates that are highly likely to be real MPTnD stars.\\

\subsection{Spatial and Kinematic Properties  }\label{3.3}

Apart from the rotational velocity ($V_\phi$), we can also discuss these MPTnD candidates from other spatial and kinematic characteristics. For spatial distributions, we focus on $R_{\rm g}$ and $Z_{max}$. In terms of kinematic properties, we investigate distributions of orbital eccentricity ($ecc$) and orbital inclination ($\theta_{L}$). It should be noted that our parent sample is based on APOGEE DR17, and the spatial coverage is incomplete due to the limitations of its survey range. Nevertheless, we can still compare the relative differences in the distributions between the MPTnD candidates and other known Galactic components. \\

\begin{figure*}[htbp]
	\centering
	\includegraphics[width=1.00\textwidth]{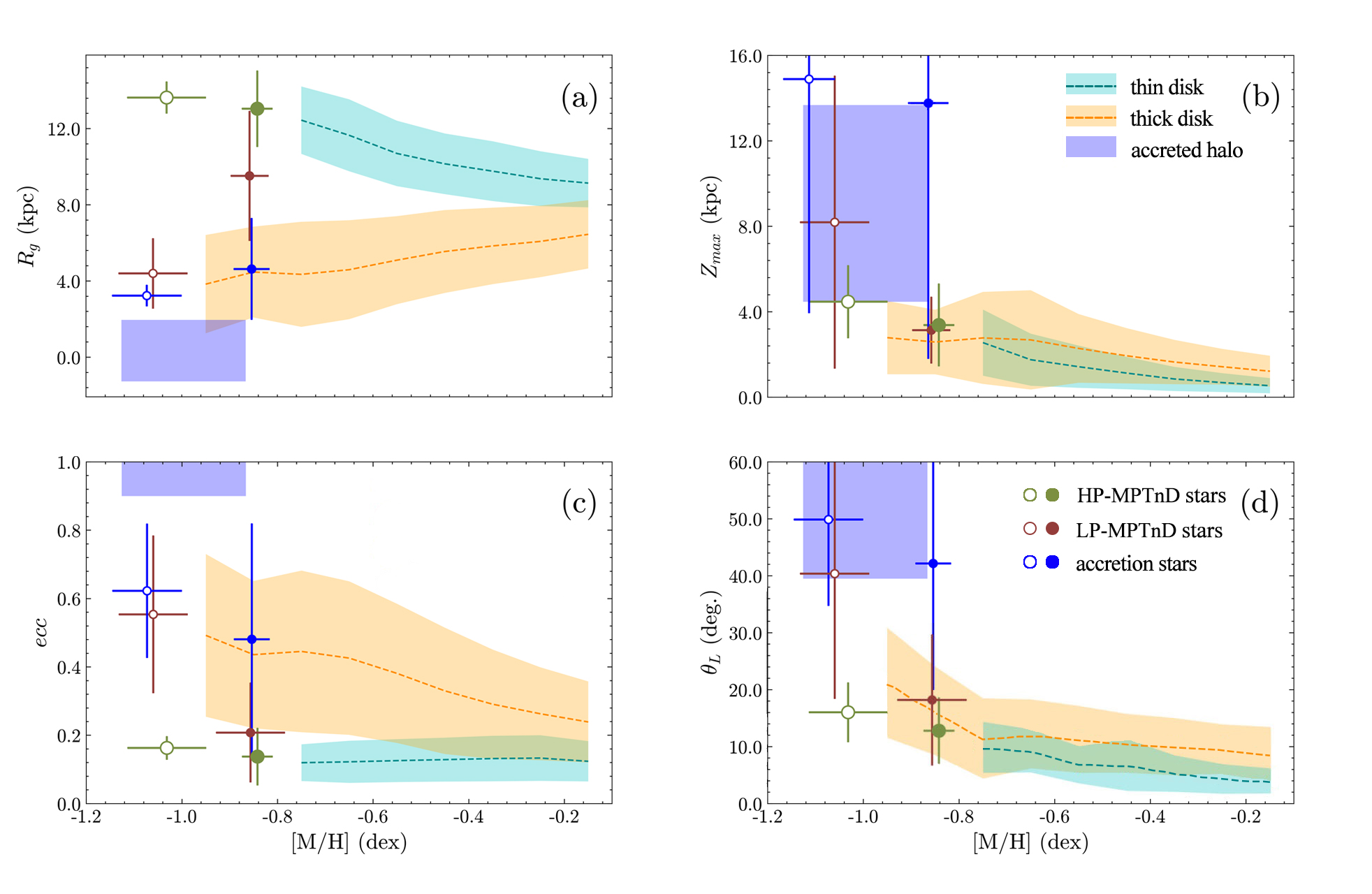}
	\caption{The spatial ($R_{\rm g}$ and $Z_{max}$) and dynamical distributions ($ecc$ and $\theta_{L}$) of Galactic components. The orange and cyan lines with shadow represent the standard deviations of corresponding parameters as functions of metallicity for the thick and thin disk, respectively. The purple rectangle represent the typical distribution ([16\%, 84\%]) of accreted halo stars (in the blue box of Fig.~\ref{Fig:2}). The green, brown and blue symbols with errorbars represent HP-MPTnD, LP-MPTnD stars and accretion stars respectively, with solid circles for [M/H] $>$ -0.95 dex and hollow circles for [M/H] $<$ -0.95 dex.  
    }

 \label{fig:cdfRZ}
\end{figure*}

Firstly, examining the HP-MPTnD stars (green symbols in all panels of Fig.~\ref{fig:cdfRZ}), it is evident that they constitute a natural extension of the thin disk (cyan dashed line with shadow) towards the more metal-poor region in all four distributions ($R_{\rm g}$, $Z_{\rm max}$, $ecc$, and $\theta_L$). Slight deviations from the thin disk and differences between two subsamples of HP-MPTnD stars ([M/H]$>$-0.95 dex and [M/H]$<$-0.95 dex, represented by solid and hollow circles, respectively) align with the expected correlations between [M/H] and these four parameters. Specifically, more metal-poor stars having slightly larger $R_{\rm g}$, $Z_{\rm max}$ and $\theta_L$ values, but maintaining relatively round orbits.  \\

 The situation of LP-MPTnD stars (brown symbols) is somewhat ambiguous. As the higher-metallicity subsample lies between extensions of the thin and thick disks, while the lower-metallicity subsample appears to follow the accreted halo. It is clear that this class of MPTnD candidates is significantly contaminated by halo or thick disk populations. \\

Compared to HP-MPTnD and LP-MPTnD stars, accretion stars are definitely closer to the accreted halo, despite differences in $R_{\rm g}$ and $ecc$. Given the diffuse distribution and larger uncertainties of accreted halo stars, these differences are within an acceptable range. Certainly, these discrepancies may also indicate that accreted halo stars with large $R_{\rm g}$ and with more circular orbits are related.\\

In all, the spatial distribution and kinematic properties further corroborate the three classification of the MPTnD candidates and reaffirms that HP-MPTnD stars are well defined metal-poorer thin disk stars. \\

\begin{figure}[htbp]
	\centering
	\includegraphics[width=0.5\textwidth]{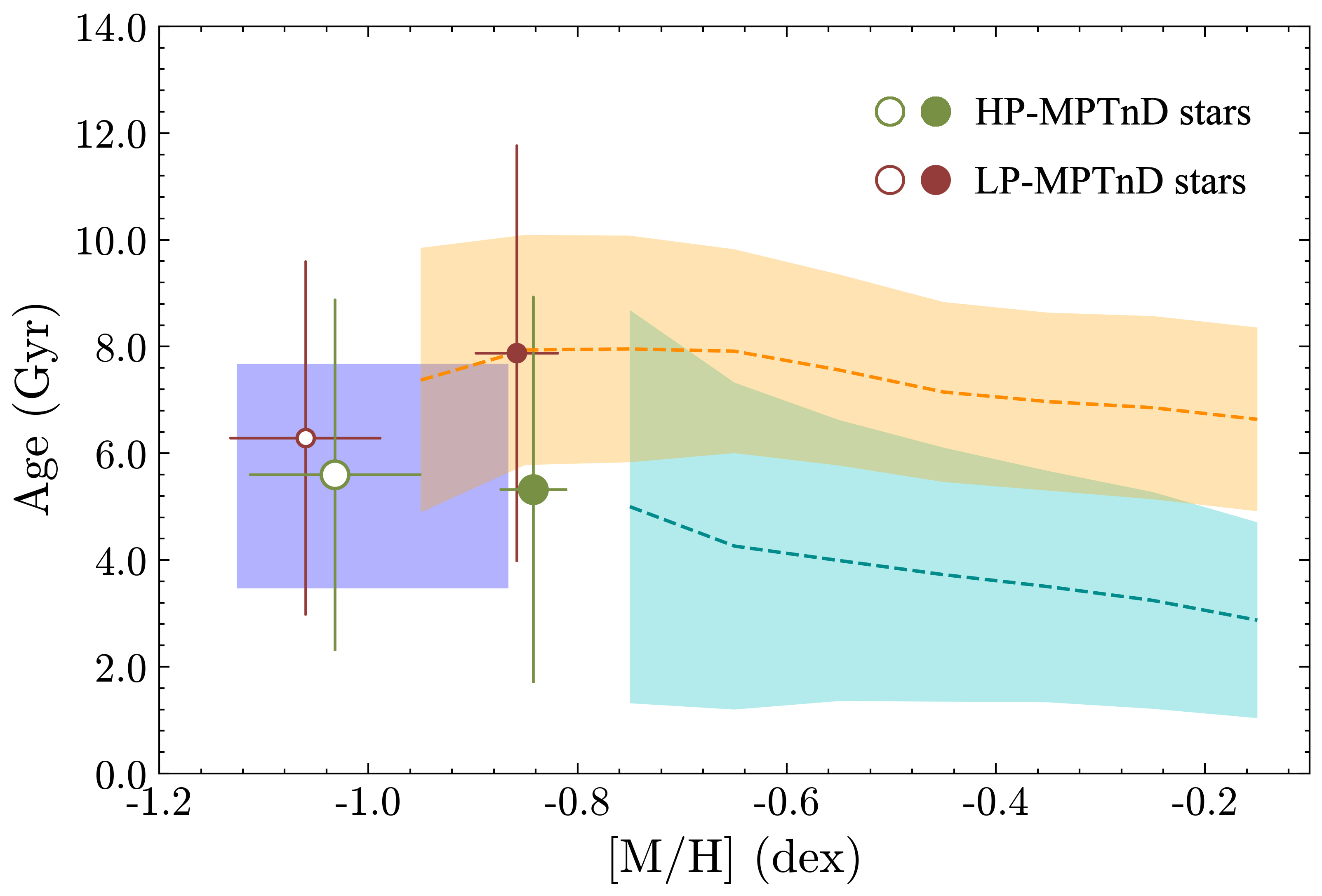}
	\caption{The age distributions for the galactic components. The symbols are the same as in Fig.~\ref{fig:cdfRZ}.}
	\label{fig:age1}
\end{figure}

\subsection{Age Distributions  }\label{3.4}

In this subsection, we use the age measurements from \cite{2018MNRAS.481.4093S}. We reject stars with age uncertainties larger than 1.5 Gyr and our age estimates of sample have typical uncertainties of $\sim$ 0.08 in log(age). 
 For our MPTnD candidates, only a subset has age measurements, all of which are either HP-MPTnD or LP-MPTnD stars, with most having [M/H]$>$ -0.95 dex (see numbers in parentheses in Table 1). Nevertheless, the shortage of data does not effect on measurements of median ages of these two subsamples. The circular symbols in Fig.~\ref{fig:age1} represent the median metallicity and age of each subsample, with error bars indicating their dispersions. The large dispersion in age is mainly due to their uncertainties rather than the intrinsic dispersion. \\

The HP-MPTnD subsamples align with the age-metallicity trend of thin disk stars, suggesting they formed in the early stage of the thin disk formation and slightly after the formation of the most metal-rich thick disk stars. Comparing with stars of similar metallicity, they are about 2.5 Gyr younger than their thick disk counterparts.\\

For LP-MPTnD stars, the age of more metal-rich subsample (solid brown symbol) is similar to that of the thick disk. While it is more dispersed, indicating a mixed population of thin disk, thick disk, and accreted halo stars. On the other hand, the more metal-poor subsample (hollow brown circle) has younger age, which is consistent with an accreted halo origin.\\

\section{Implications on formation of the galactic disc} \label{sec:4}

Through the analysis of Section~\ref{sec:3}, we confirmed the existence of a population of more metal-poor stars within the low-$\alpha$ sequence, including at least 56 HP-MPTnD stars between -1.2 dex $<$ [M/H] $<$ -0.8 dex.
This result is similar to the recent work of \cite{2024A&A...685A.151F}, who claims that \emph{the chemo-dynamically defined thin disk begins to appear at metallicities between -1 and -0.7 dex.} \\

The existence of these more metal-poor thin disk stars, which at least down to [M/H] $\sim$ -0.95 dex, can serve as a diagnostic tool to test different models of Galactic disk formation. In the following, we will briefly discuss this issue in the context of the two most prevailing models: the continuous-accretion model and the two-infall model.\\

The continuous-accretion model posits that the thick disk and thin disk are two components of a single, continuously formed Galactic disk \citep[e.g.,][]{2021MNRAS.507.5882S,2021ApJS..254....2P}. Within this framework, typical metal-poor stars ([M/H] $<$ -0.8) formed during the early stages of disk formation. A subset of these stars, unaffected by subsequent disk heating processes, may retain thin disk-like kinematics. However, the short timescale of the early star formation resulted in elevated $\alpha$-enhancement ([$\alpha$/M] $>$ 0.25 dex), which is clearly inconsistent with the observed low-$\alpha$ characteristics of HP-MPTnD stars. Alternatively, low-$\alpha$ stars primarily originate from the secular star formation process in the later stage. Despite slower metal enrichment in the outer disk leading to lower stellar metallicities, the formation of very metal-poor stars remains challenging. For instance, according to \cite{2021MNRAS.507.5882S}, in the outermost regions of the disk ($R_{\rm g} \sim 12-14 $ kpc), which is comparable to the $R_{\rm g}$ of our HP-MPTnD sample, low-$\alpha$ sequence stars are predicted to have [M/H] $\sim$ -0.5 dex. This value is significantly higher than the well-established metallicity floor of -0.95 dex for our HP-MPTnD sample. Even considering observational uncertainties. Thus, the likelihood of observing such a low-$\alpha$ metal-poor population in the outer disk region is exceedingly small. In other words, the existence of HP-MPTnD stars seems to have an irreconcilable contradiction with the predictions of the continuous-accretion model.\\

In contrast, the two-infall model offers a more flexible framework to address this issue. Within this model, the early gas infall triggered the first starburst and contributed to the formation of the high-$\alpha$ sequence (thick disk stars). The subsequent secular star formation of low-$\alpha$ stars (outer thin disk stars) was caused by second major gas infall \citep{Chiappini97,Lian20}. The late accretion of gas diluted the local Milky Way's gas, which had been enriched by the first infall. Then, the metallicity of stars formed in the early phase of the second gas infall is completely determined by the mixing ratio of two kinds of gas and their individual metallicities. Thus, these stars can exhibit a broad range of metallicities. \\

Nevertheless, the existence of HP-MPTnD stars can provide significant constraints on the two-infall model. Firstly, it offers insights into the timing of the second infall, which coincides with the birth of MPTnD stars. Given the potential systematic biases from different age measurement methods, we confine our discussion within the sample  of \cite{2018MNRAS.481.4093S}, to compare the relative timings of the second infall with other disk components. The second infall event occurred approximately 5.5 Gyr ago.  Referring to Fig.~\ref{fig:age1} of this paper and Fig. 7 of HS22, we can infer that the second major gas infall began shortly after the formation of the canonical thick disk,  but slightly before the formation of inner part of the thin disk. This might suggest that the second infall gas cooling down more efficiently in the outer region of Galactic disk, leading to earlier star formation.\\

Secondly, MPTnD stars impose a stringent constraint on the metallicity of the infalling gas. The infalling gas must be sufficiently metal-poor, otherwise mixing with the local gas would easily lead to a more metal-rich thin disk tail. For example, with only 11\% of the local gas having a metallicity of [M/H]=0 dex, it is possible to enrich the pristine infalling gas to [M/H]=-0.95 dex. That means a major gas infall with large mass and extremely metal-poor is required. Alternatively, if the infalling gas comes from accreted dwarf galaxies, the metallicity of MPTnD stars-assuming they are entirely formed from this infalling gas-imposes an upper mass limit on dwarf galaxies. Since there is liner correlation between the mass and gas metallicity of dwarf galaxies (Equation (7) in \citealt{2005ApJ...635..260S}), the more massive dwarf galaxy, the more metal-rich of its remaining gas. Then, the upper limit of gas metallicity results in an upper limit on galaxy mass. Taking the [O/H] abundance of our HP-MPTnD sample ($\sim$ -0.8 dex) that is derived from APOGEE DR17, we estimate an upper limit of approximately 10$^6 M_{\odot}$ for the mass of individual accreted dwarf galaxies. Such a mass is significantly smaller than that of the Milky Way's thin disk. Therefore, it implies that the Milky Way must have experienced a late-time infall of large amount of relatively pristine gas.\\

While the qualitative analysis above cannot entirely rule out the continuous model, it suggests that the more metal-poor the stars are at the low-$\alpha$ end, the greater the likelihood that the galactic disk formed through two or more major gas infall events.

\section{Conclusion}\label{sec:Conclusion}

This work aimed to explore the metal-poor tail of the thin disk and shed light on the formation of the Galactic disk. Through our chemical and kinematic analyses, we demonstrated that thin disk stars can exhibit metallicities as low as -0.8 to -1.2 dex, significantly lower than the previously established lower limit of -0.7 dex. \\

Based on the chemical and kinematic data from APOGEE DR17 and Gaia DR3, we firstly identified a set of the metal-poor thin disk (MPTnD) candidates with lower $\alpha$-enhancement and higher rotational velocity through an examination of the [$\alpha$/M]- $V_{\phi}$ plane. Using this sample, we confirm the claim of some previous works that the lower metallicity limit of the thin disk should be lower than -0.95 dex.\\

We further refined this sample by excluding contaminated accreted-halo stars and suspected thick disk stars using specific elemental abundance ratios ([Mg/Mn], [Al/Fe], and [C+N/Fe]). This process resulted in the confirmation of 56 high-possibility MPTnD stars. Subsequent analyses of spatial ($R_{\rm g}$, $Z_{\rm max}$) and kinematic ($ecc$, $\theta_L$) distributions confirmed that these HP-MPTnD stars belong to the metal-poor tail of the Milky Way's thin disk.\\

The existence of these very metal-poor thin disk stars offers valuable constraints on the Milky Way's disk formation scenario. The observed low metallicity limit of the thin disk supports the two-infall model, suggesting that the second major gas infall event occurred slightly later than the formation of the thick disk.\\

This work was supported by the National Key R\&D Program of China No. 2019YFA0405501; the National Natural Science Foundation of China (NSFC) under grants 12273091, U2031139 and 12173013; the Science and Technology Commission of Shanghai Municipality (Grant No.22dz1202400);  the science research grants from the China Manned Space Project with NO. CMS-CSST-2021-A08; the Postdoctoral Fellowship Program of CPSF under Grant Number GZC20240371; Science Foundation of Hebei Normal University (No. L2024B55). This work was also sponsored by Program of Shanghai Academic/Technology Research Leader. The work of EG is supported by National Natural Science Foundation of China (NSFC) programme nos. 11988101 under young talent project QN2023061004L. This work has made use of data from the European Space Agency (ESA) mission Gaia (https://www.cosmos.esa.int/gaia), processed by the Gaia Data Processing and Analysis Consortium (DPAC; https://www.cosmos.esa.int/web/gaia/dpac/consortium). Funding for the DPAC has been provided by national institutions, in particular the institutions participating in the Gaia Multilateral Agreement. The stellar parameters, abundances, RVs from APOGEE DR17 were derived from the allStar files available at https://www.sdss.org/dr17/. Funding for the Sloan Digital Sky Survey IV has been provided by the Alfred P. Sloan Foundation, the U.S. Department of Energy Office of Science, and the Participating Institutions. \\

\software{Numpy \citep{2011CSE....13b..22V}, Scipy \citep{2007CSE.....9c..10O}, Matplotlib \citep{2007CSE.....9...90H}.}

\bibliography{ref}{}
\bibliographystyle{aasjournal}

\end{CJK*}	
\end{document}